\documentclass[a4paper,11pt]{article}
\usepackage{pos}
\usepackage{amsmath,leftidx}
\usepackage{dsfont}
\usepackage{bbm}
\usepackage{cleveref}
\usepackage{subcaption}
\usepackage[utf8]{inputenc}
\usepackage{mathrsfs}
\usepackage{amsfonts}
\usepackage{wrapfig}
\usepackage{enumitem}
\usepackage{todonotes}

\usepackage{defs}

\title{$m_B$ and $f_{B^{(\star)}}$ in $2+1$ flavour QCD from a combination of continuum limit static and relativistic results}

\author[ma,mz,g]{Alessandro Conigli}
\author[n]{Julien Frison}
\author[d]{Patrick Fritzsch}
\author[mar]{Antoine G\'erardin}
\author[mu]{Jochen Heitger}
\author[ma]{Gregorio Herdoiza}
\author[c,mz,g]{Simon Kuberski}
\author[ma]{Carlos Pena}
\author[n]{Hubert Simma}
\author[n,h]{Rainer Sommer}

\affiliation[n]{John von Neumann-Institut f{\"u}r Computing NIC, Deutsches Elektronen-Synchrotron DESY,\\
	Platanenallee 6, 15738 Zeuthen, Germany}
\affiliation[h]{Institut f{\"u}r Physik, Humboldt-Universit{\"a}t zu Berlin\\
	Newtonstr. 15, 12489 Berlin, Germany}

\affiliation[ma]{Instituto de F\'{\i}sica Te\'orica UAM-CSIC and Dpto. de F\'{\i}sica Te\'orica\\
	C/~Nicol\'as Cabrera 13-15, Universidad Aut\'onoma de Madrid, Cantoblanco E-28049 Madrid, Spain}

\affiliation[mu]{Universit\"at M\"unster, Institut f\"ur Theoretische Physik,\\
	Wilhelm-Klemm-Stra{\ss}e 9, 48149 M\"unster, Germany}

\affiliation[mar]{Aix-Marseille Universit\'e, Universit\'e de Toulon, CNRS, CPT, Marseille, France
}

\affiliation[d]{School of Mathematics, Trinity College Dublin, Dublin 2, Ireland}

\affiliation[c]{Theoretical Physics Department, CERN, 1211 Geneva 23, Switzerland}

\affiliation[mz]{Helmholtz Institute Mainz, Johannes Gutenberg University, Mainz, Germany}
\affiliation[g]{GSI Helmholtz Centre for Heavy Ion Research, Darmstadt, Germany}

\emailAdd{aconigli@uni-mainz.de}

\abstract{
We present preliminary results for B-physics from a combination of non-perturbative results in the static limit with relativistic computations satisfying $am_{\mathrm{heavy}}\ll 1$. Relativistic measurements are carried out at the physical b-quark mass using the Schr\"{o}dinger Functional in a $0.5 \ \mathrm{fm}$ box. They are connected to large volume observables through step scaling functions that trace the mass dependence between the physical charm region and the static limit, such that B-physics results can be obtained by interpolation; the procedure is designed to exactly cancel the troublesome $\alpha_s(m_{\mathrm{heavy}})^{n+\gamma}$ corrections to large mass scaling. Large volume computations for both static and relativistic quantities use CLS $N_f=2+1$ ensembles at $m_u=m_d=m_s$, and with five values of the lattice spacing down to $0.039$ fm.
Our preliminary results for the b-quark mass and leptonic decay constants have competitive uncertainties, which are furthermore dominated by statistics, allowing for substantial future improvement. Here we focus on numerical results, while the underlying strategy is discussed  in a companion contribution \cite{RainerPos}.
}

\FullConference{The 40th International Symposium on Lattice Field Theory (Lattice 2023)\\
July 31st - August 4th, 2023\\
Fermi National Accelerator Laboratory\\}

\begin{document}
\maketitle

\section{Introduction}
Fundamental processes involving heavy quarks are crucial to explore indirect searches of new physics beyond the Standard Model (SM).  In this context, a precise non-perturbative determination of  B-physics observables is essential to probe new phenomena that may manifest themselves as subtle deviations from the theoretical predictions.

We have further developed a methodology first proposed in \cite{ Guazzini:2007ja}, where a static computation $m_h\to\infty$ was combined  with a relativistic one  $m_h <m_b$  in order to reach the physical b-quark scale by interpolation while controlling cutoff effects in each step of the computation.  Our approach, detailed  in \cite{RainerPos}, consists in building suitable quantities with a continuum limit and  a simple behavior in $1/m_h$ in order to perform the interpolation between the two sets of results.  The design principle is a complete cancellation of renormalisation and matching factors in the static theory. The latter  diverge  in the static limit, thus posing serious difficulties to control systematic uncertainties.

Our strategy requires a set of finite volume ensembles, from $L_1=2L_0\approx 0.5 \ \mathrm{fm}$  where we reach the relativistic b-quark scale, to $L_2\approx 1.0 \ \mathrm{fm}$, where we perform both static and relativistic measurements. We cover a range of heavy quark masses that starts below the charm region and extends up to or above the bottom quark mass, while the light quark masses are set to zero in finite volume.  In  $L_2$, as we double the box size and the lattice spacing, discretisation effects increase significantly above $m_b/2$.   Eventually we make contact with Nature by connecting to the large volume $SU(3)$ symmetric point $m_u=m_d=m_s$ through CLS ensembles~\cite{Bruno:2014jqa, Bruno:2016plf, Mohler:2017wnb, Mohler:2020txx}.  Different volumes are connected by step scaling \cite{Heitger:2001ch,deDivitiis:2003iy,Guazzini:2007ja}.

\section{b-quark mass}\label{sec:bquarkmass}
We extract the b-quark mass from the step scaling chain \cite{RainerPos}
\begin{equation}
	m_b^{\mathrm{RGI}} = \frac{1}{L_{\mathrm{ref}}} \frac{y_B - \rho_m(u_2, y_B) - \sigma_m(u_1, y_2)}{\pi_m(u_1, y_1)},
\end{equation}
where we define the Step Scaling Functions (SSF)
\begin{equation}\label{eq:mb_master_eq}
	\sigma_m(u_1, y_2) = L_{\mathrm{ref}}[m_H(L_2) - m_H(L_1)], \qquad 
	\rho_m(u_1, y_2) = L_{\mathrm{ref}}[m_B - m_H(L_2)],
\end{equation}
made dimensionless in units of a length-scale $L_{\mathrm{ref}}$. We specify the different volumes of size $L_i^4$ in terms of the  running couplings $u_i = \bar{g}^2(L_i)$ of \cite{DallaBrida:2016kgh,Fritzsch:2018yag},
\begin{equation}
	u_0=3.949,\; u_1=5.862(27),\;  u_2= 11.252(83)\,,
\end{equation}
while
 the heavy-light meson masses $y_i = L_{\mathrm{ref}}m_H(L_i)$ in units of $L_{\mathrm{ref}}$ are used as proxies for the b-quark mass\footnote{Here we refer to $m_H(L)$ as finite volume heavy light pseudo-scalar masses, defined as in \cite{Guazzini:2007ja}. They have the property $\lim_{L\to\infty}m_H(L) = m_H$ with $m_H$ the true particle mass. }.  They are determined recursively from the large volume physical input according to
 \begin{equation}
 	y_B\equiv L_{\mathrm{ref}}m_{\overline{B}}, \qquad y_2 = y_B - \rho_m(u_2, y_B), \qquad y_1 = y_2 - \sigma_m(u_1, y_2),
 \end{equation}
where $m_{\overline{B}} = \frac{2}{3} m_B + \frac{1}{3} m_{B_s} = 5308.5(2) \ \mathrm{MeV}$ denotes the flavour averaged combination  of $B$ and $B_s$ meson masses \cite{ParticleDataGroup:2020ssz} used to fix the physical b-quark mass. This is the natural choice at the symmetric point ($M_\pi = M_K \approx 415 \ \mathrm{MeV}$ \cite{Bruno:2017gxd}).
 Finally, $\pi_m(u_1, y_1)$  in Eq.~(\ref{eq:mb_master_eq}) is the ratio  
\begin{equation}
		\pi_m(u_1, y_1) = \frac{m_H(L_1)}{m_b^{\mathrm{RGI}}} = \frac{y_1}{L_{\mathrm{ref}}m_b^{\mathrm{RGI}}},
		\label{eq:pi_m_def}
\end{equation}
 used to convert pseudo-scalar masses to RGI quark masses,
 \begin{equation}
 	m_h^{\mathrm{RGI}} = h(L_0) \frac{Z(g_0^2) Z_A(g_0^2)}{Z_P(g_0^2, L_0)}(1 + b_m(g_0^2) am_{q,h})m_{q,h},
 \end{equation}
with $m_{q,h}$ the bare subtracted quark mass. The renormalisation constants $Z_A$, $Z_P$, $Z$ and $b_m$ have been determined specifically   for our bare coupling range~\cite{Fritzsch:2018yag, Kuberski:2020rjq},  while the non-perturbative RG running factor $h(L_0) =1.4744(85)$ has been computed using the results of \cite{Campos:2018ahf}.

We now move to the numerical results for the SSFs. In our practical implementation we set $L_{\mathrm{ref}} = 4L_0 = L_2$.

\subsection{$L_2$ to $L_{\mathrm{CLS}}$ step scaling functions}
This first step connects the finite volume  $L_2$  with vanishing sea quark masses to  the large volume CLS ensembles at the symmetric point. 
Bare couplings were tuned such that  $\bar g^2(L_2)=11.27$ (a preliminary value at the start of the project) within a sufficient precision $\Delta \bar g^2=0.1$ for $L_2/a= 12, \ldots, 32$. This results in $\beta\in [3.4, 3.97]$  overlapping well with  CLS ensembles. Short interpolations  of $L/a=f(\tilde g_0^2)$ to the CLS improved bare parameters $\tilde{g}_0^2$ (see \cite{Luscher:1996sc, Bhattacharya:2005rb} for the definition) are required to match the  two sets of data. Next finite volume and infinite volume heavy-light pseudo-scalar masses are computed at a suitably chosen set of bare heavy quark masses $a \tilde m_{\mathrm{q},h}$. For each (non-integer) resolution $a/L$  all heavy-light quantities are then 
interpolated to common improved bare parameters 
$\tilde{g}_0^2,a\tilde m_{\mathrm{q},i}$ and these are  interpolated
again to a set of $\{y^\mathrm{targ}\}$ (defined in large volume). For each $y^\mathrm{targ}$ we then have finite volume and infinite volume
masses for the same improved bare parameters at each resolution  $a/L_2$.  

This  yields the relativistic SSFs at finite lattice spacing,
\begin{equation}\label{eq:rhom_rel_finite_a}
	R_m(u_2, y^\mathrm{targ}, a/L_2) = L_2[ (m_H)_{\mathcal{S}_m} - (m(L_2))_{\mathcal{S}_0}]\,. 
\end{equation}
The static ones are obtained in the same way but do not need an interpolation to a fixed $y^\mathrm{targ}$.	
Here   $\mathcal{S}_0$ and $\mathcal{S}_m$ denote the massless and massive Lines of Constant Physics (LCP)  defined in \cite{RainerPos} and connected by the common improved bare coupling $\tilde g_0^2$  as just explained.

\begin{figure}[!h]
	\centering
	\includegraphics[scale=0.455]{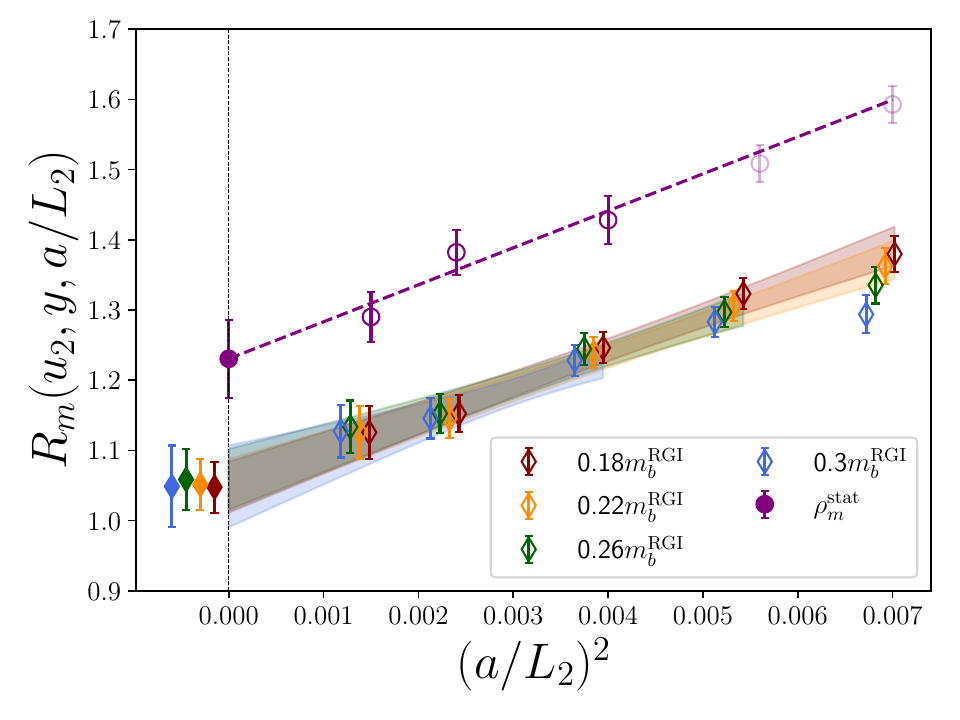}
	\includegraphics[scale=0.465]{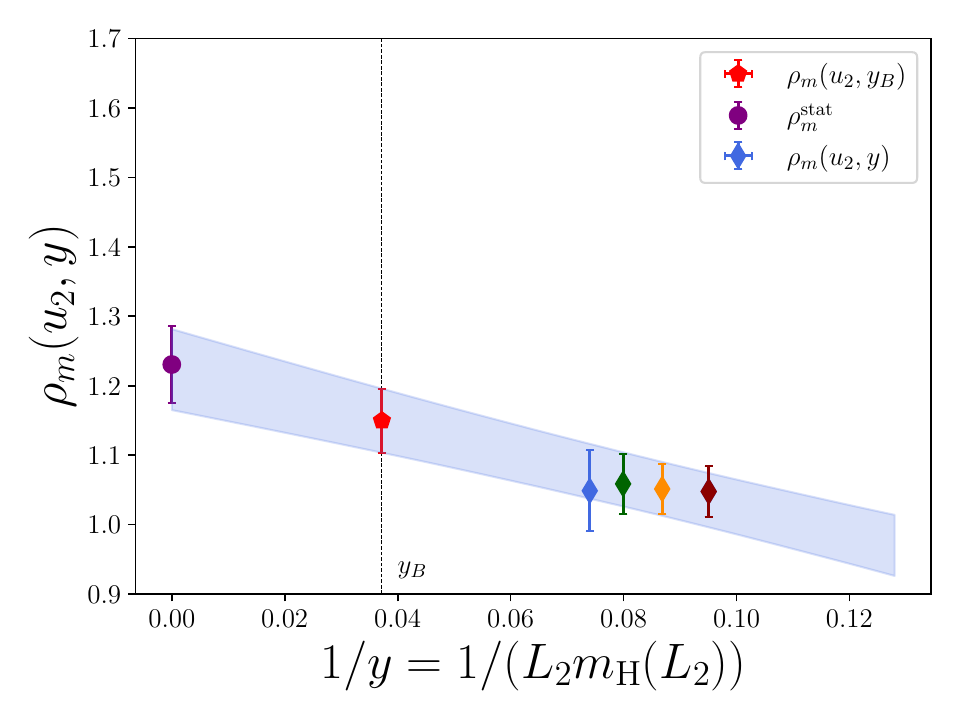}
	\caption{\textit{Left:} continuum limit extrapolation for the relativistic (diamonds) and static (circles) SSF connecting the volume $L_2\to \mathrm{CLS}$. \textit{Right:} interpolation between the  static and relativistic continuum SSF  to the target b-quark scale $y_B$.}
	\label{fig:ssf_rel_L2_CLS}
\end{figure}

 The extrapolation to $\rho(u,y)=R_m(u,y,0)$ is performed with the fit ansatz
\begin{equation}\label{eq:rho_m_cont_fit}
	R_m = p_0 + p_1 (a/L_2)^2, \qquad 	R_m^{\mathrm{stat}} = s_0 + s_1 (a/L_2)^2.
\end{equation}
Cutoff effects
are compatible with a linear dependence on $a^2$, cf. Fig.~\ref{fig:ssf_rel_L2_CLS}. For the relativistic $R_m(u_2,  y^\mathrm{targ}_i, a/L_2)$ we include 
only data  with $am_h^{\mathrm{RGI}}<0.8$
in the fit  and in the static theory we drop the two
coarsest lattice spacings.  The improvement coefficient $c_A^{\mathrm{stat}}$ is needed for O($a$) improvement of $R_m^{\mathrm{stat}}$. However, $c_A^{\mathrm{stat}}$ is not yet known  for the L\"{u}scher-Weisz gauge action.  We currently estimate it by the 1-loop  $c_A^{\mathrm{stat}}$ of the Wilson gauge action \cite{Grimbach:2008uy} and assign a $200\%$ error.  

We then interpolate the continuum results for $\rho_m(u_2, y^\mathrm{targ})$ and $\rho_m^{\mathrm{stat}}(u_2)$  to the physical point $y=y_B=26.87(18)$ through the functional form
\begin{equation}
	\rho_m = t_0 + t_1 y^{-1} + t_2 y^{-2},
	\label{eq:mb_interp_static_rel}
\end{equation}
where $t_i$ are fit parameters.  The interpolation is shown on the right plot of Fig.~\ref{fig:ssf_rel_L2_CLS}. In this case, the quadratic term  is not needed in the fit and we set $t_2=0$.  At the physical point we deduce
\begin{equation}
	\rho_m(u_2, y_B) = 1.152(45)\,,\quad\to\quad y_2=25.72(19).
\end{equation}

\subsection{$L_1$ to $2L_1$ step scaling functions }
In the second step of the computation we determine the SSF $\sigma_m(u_1, y_2)$  connecting the finite volumes $L_1$ and $L_2=2L_1$, generated at common values of the  inverse coupling in the range $\beta\in [3.65,4.25]$. The relativistic measurements were directly performed at equal values of all bare parameters, including the heavy quark masses from the charm region up to the bottom quark mass. Therefore, interpolations are only needed to common $\{y^\mathrm{targ}\}$ but not in $am_{q.h}$. 

\begin{figure}
	\centering
	\includegraphics[scale=0.46]{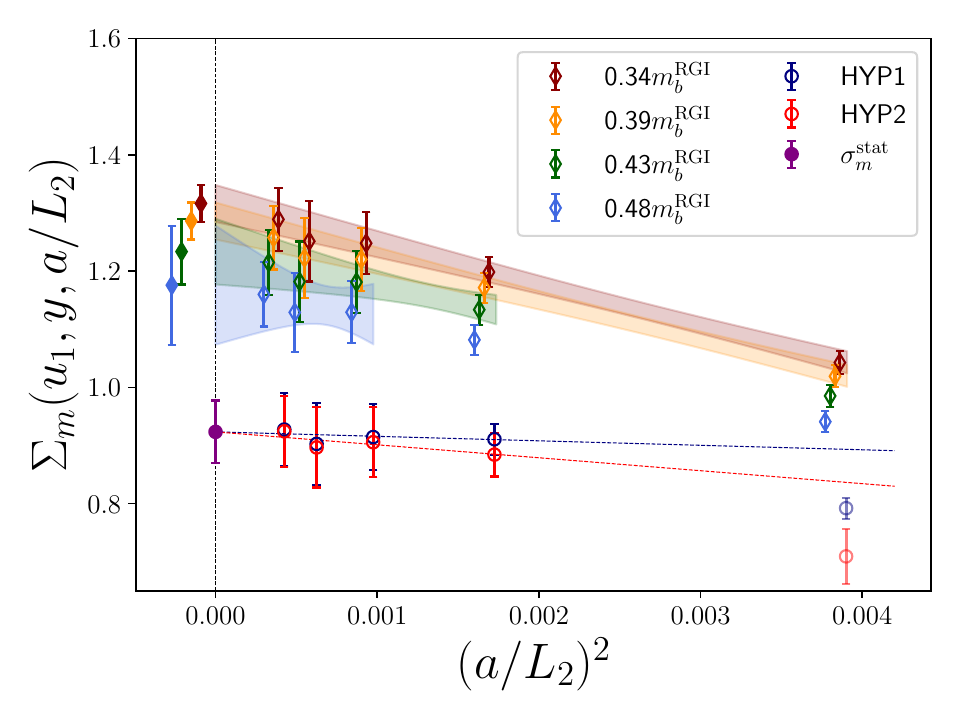}
	\includegraphics[scale=0.46]{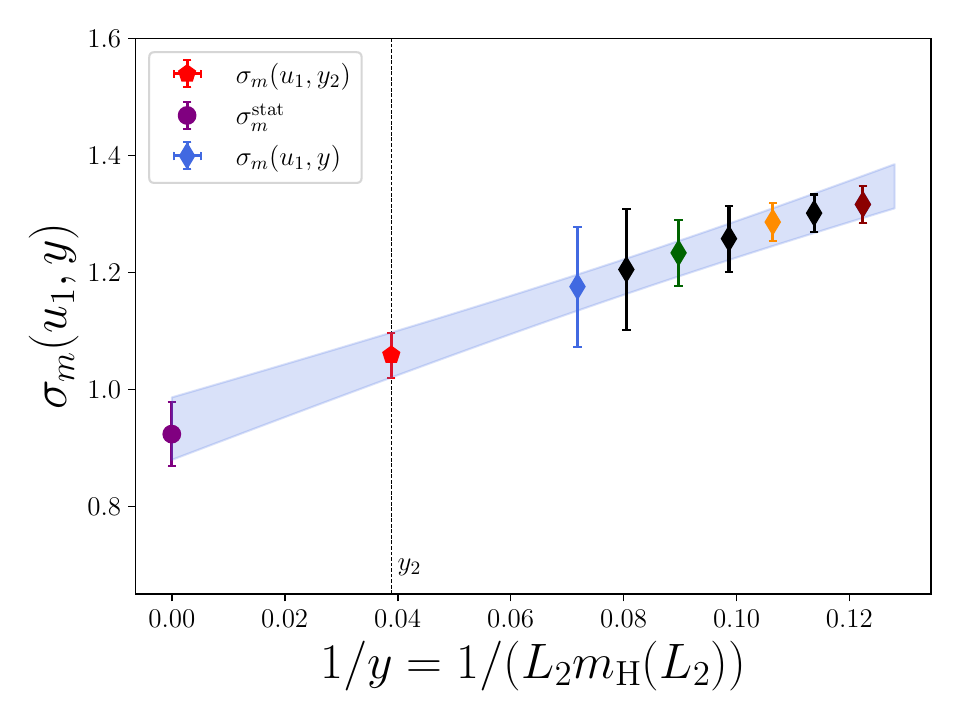}
	\caption{\textit{Left}:  continuum limit extrapolation for the relativistic (diamonds) and static (circles) SSF connecting the volume $L_1\to L_2=2L_1$. In the relativistic regime we show only a subset of the available heavy masses to improve readability. In the static theory we make use of two different actions $\mathrm{HYP}1$ and $\mathrm{HYP}2$ to constrain the continuum limit. \textit{Right:} interpolation between static and relativistic continuum SSF to the physical point $y_2$.  The colour code is the same in both plots. 
	}
	\label{fig:ssf_L1_to_L2}
\end{figure}

In analogy with Eq.~(\ref{eq:rhom_rel_finite_a}) we define the finite-$a$ SSFs, 
\begin{equation}
		\Sigma_m(u_1, y, a/L_2) = L_2[ m_H(L_2) - m(L_1)]\big|_{\mathcal{S}_0}, \qquad  		\Sigma_m^{\mathrm{stat}}(u_1, a/L_2) = L_2[ E^{\mathrm{stat}}(L_2) - E^{\mathrm{stat}}(L_1)]\big|_{\mathcal{S}_0},
\end{equation}
used to extract the continuum $\sigma_m(u_1, y)$ and $\sigma_m^{\mathrm{stat}}(u_1)$. We parametrise the lattice spacing dependence as in   Eq.~(\ref{eq:rho_m_cont_fit}). Fig.~\ref{fig:ssf_L1_to_L2}  suggests a good control over cutoff effects. Nonetheless, we impose the same cuts on the data entering the fits as before. It is worth noticing that in the static sector we employ two different static actions named HYP1 and HYP2 \cite{Hasenfratz:2001hp,DellaMorte:2013ega} to further constrain the continuum limit. 
 
The physical point is then reached by interpolating $\sigma_m(u_1, y)$  and  $\sigma_m^{\mathrm{stat}}(u_1)$ to $y=y_2=25.72(19)$, c.f. Fig.~\ref{fig:ssf_L1_to_L2}.  We quote as  result
\begin{equation}
	\sigma_m(u_1, y_2) = 1.058(49)\,,\quad\to\quad y_1=24.66(19).
\end{equation}

\subsection{Relativistic QCD in $L_1$}
\begin{figure}[!h] 
	\centering
	\includegraphics[scale=0.46]{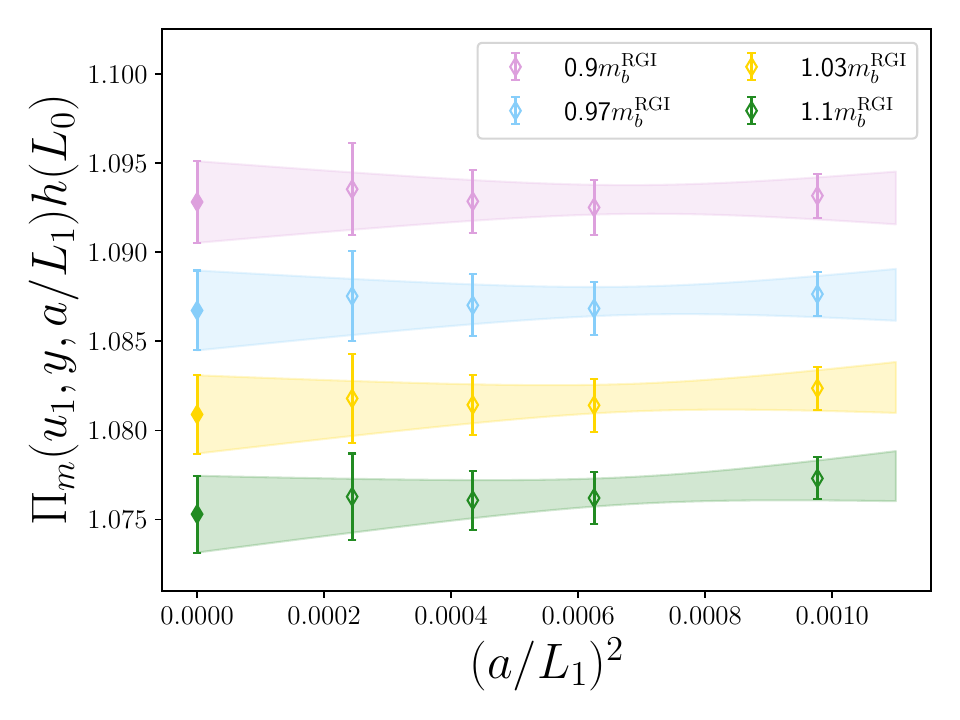}
	\includegraphics[scale=0.46]{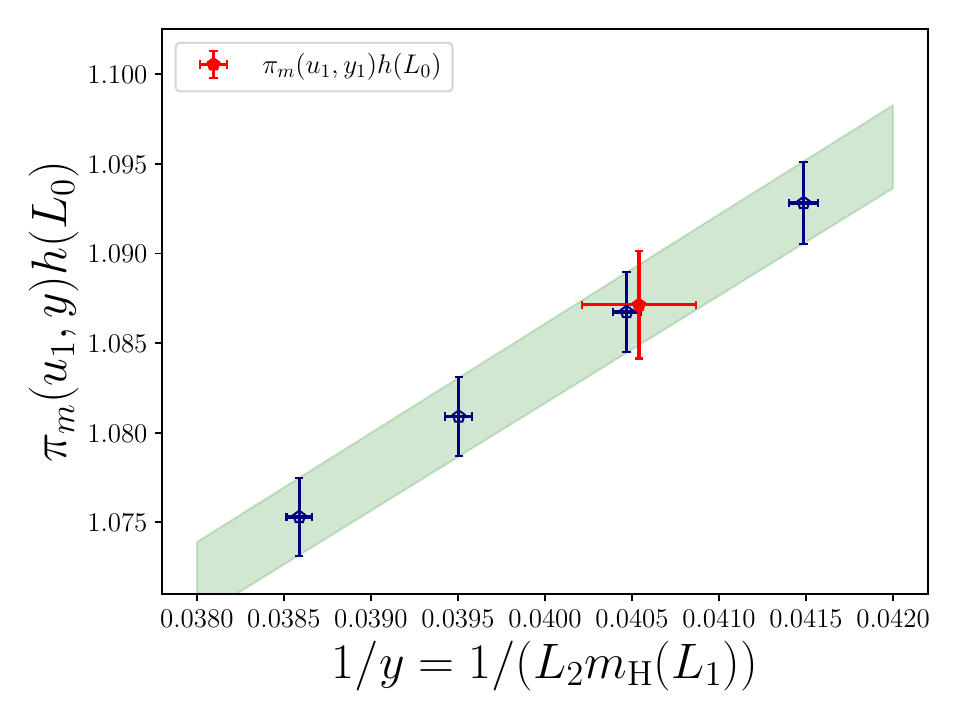}	
	\caption{\textit{Left:} continuum  extrapolation of $\Pi_m(u_1,y,a/L_1)$ at four different masses around the b-quark region in the finite volume $L_1$. A fifth available coarsest  lattice spacing is excluded from the fit. \textit{Right:} interpolation to the target b-quark scale $y_1$. }
	\label{fig:L1_rel_qcd_mb}
\end{figure}

Finally, we compute the function $\Pi_m(u_1, y_1,a/L)$, 
the finite $a$ approximant to $\pi$, Eq.~(\ref{eq:pi_m_def}), in the  volume $L_1\approx 0.5 \ \mathrm{fm}$. Here we simulate very fine lattice spacings $0.008 \ \mathrm{fm} \leq a \leq 0.02 \ \mathrm{fm}$, directly around $y_B$.
Again we assume a linear dependence on $a^2$ to parametrise the cutoff effects at quark masses between $0.9m_b$ and $1.1m_b$. The continuum  extrapolation together with the following linear interpolation in $y$ 
to the physical point $y_1$ are shown in Fig.~\ref{fig:L1_rel_qcd_mb}. We observe a remarkably weak dependence on the lattice spacing. To a significant part this is due to the use of a massive scheme for $Z$ and $b_\mathrm{m}$ \cite{Fritzschetal:tocome}.

We can now evaluate Eq.~(\ref{eq:mb_master_eq}) to quote our preliminary result
\begin{equation}
	m_b^{\mathrm{RGI}}(N_f=3) = 6.608(49) \ \mathrm{GeV} \ [0.7\%]\,,
\end{equation}
where we remind the reader that the input is the SU(3) symmetric point in the sea sector. Given experience with very small deviations to that point (see Fig. 4 in \cite{Heitger:2021apz} and Fig. 8 in \cite{ Bussone:2023kag}), we quote no 
extra uncertainty but are planning to extend the computation to the physical point in $m_\pi$ and $m_K$
in the future. 

\begin{figure}[!h] 
	\centering
	\includegraphics[scale=0.4]{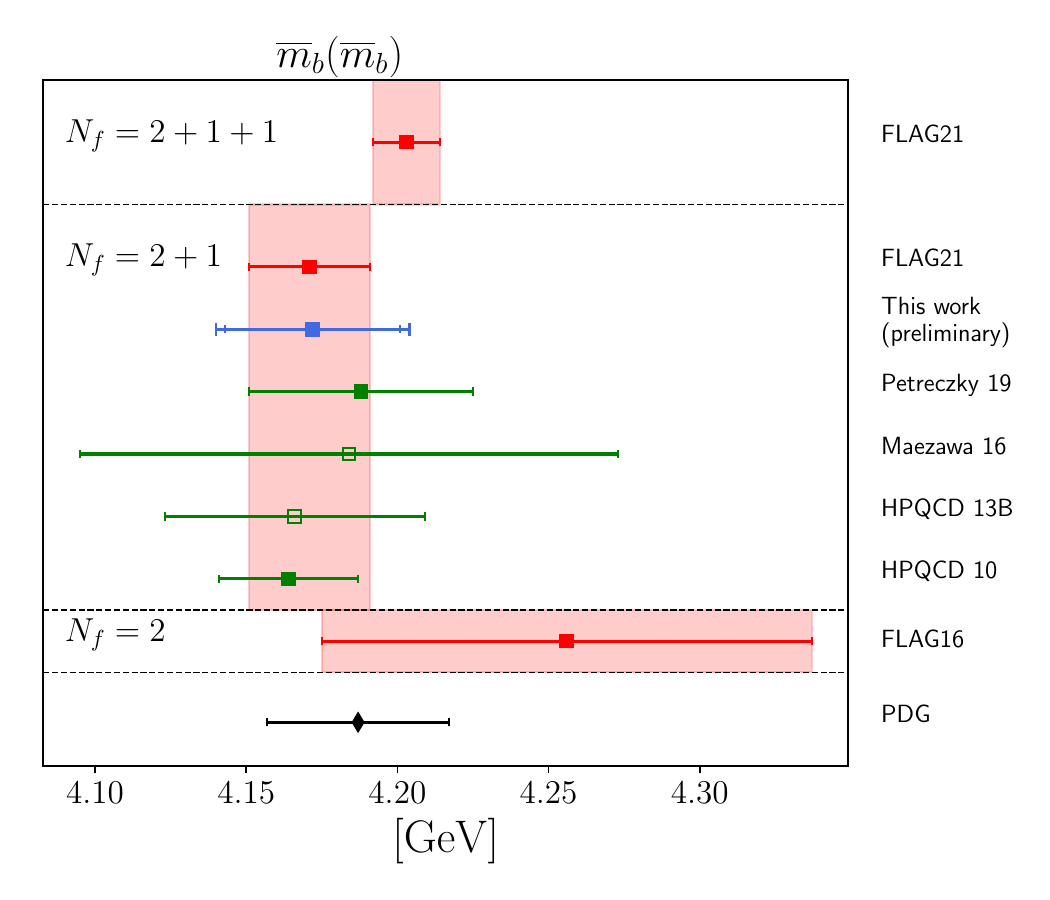}
	\includegraphics[scale=0.46]{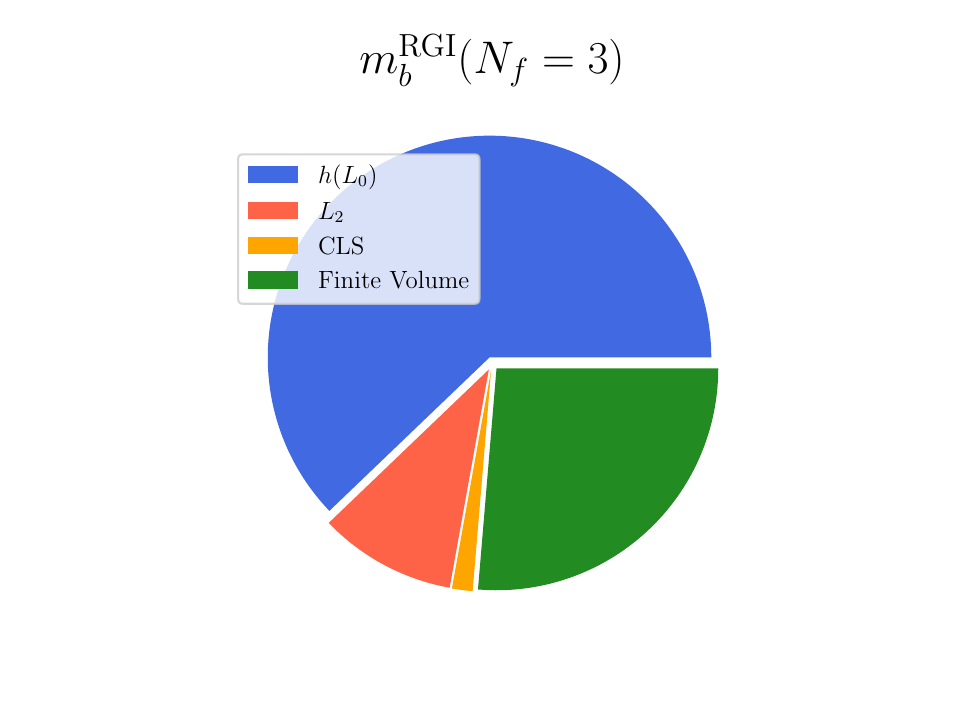}	
	\caption{ \textit{Left}: comparison of our results for $	\overline{m}_b(\overline{m}_b)$ with FLAG averages  and results from other collaborations.  For $N_f=2+1+1$  \cite{FlavourLatticeAveragingGroupFLAG:2021npn} the determinations  \cite{Hatton:2021syc, FermilabLattice:2018est, Gambino:2017vkx, ETM:2016nbo, Colquhoun:2014ica, Bussone:2014cha, Chakraborty:2014aca} contribute to the average.
		For $N_f=2+1$ \cite{FlavourLatticeAveragingGroupFLAG:2021npn} results, converted to $N_f=4$, are taken from:  HPQCD10 \cite{PhysRevD.82.034512},  HPQCD 13B \cite{PhysRevD.87.074018}, Maezawa 16 \cite{PhysRevD.94.034507}, Petreczky \cite{PhysRevD.100.034519}. Only the green filled points contribute to the FLAG average. For $N_f=2$ \cite{Aoki:2016frl} the values contributing to the average are: \cite{ETM:2013jap, Bernardoni:2013xba, ETM:2011zey}; they are of course not converted to the $N_f=4$ theory. The PDG value   \cite{ParticleDataGroup:2020ssz} is converted from their $N_f=5$ result to $N_f=4$.  \textit{Right:} contribution to the variance of $m_b^{\mathrm{RGI}}$. The dominant error source comes from the non-perturbative running factor $h(L_0)$  \cite{Campos:2018ahf}, followed by the finite volume measurements and the reference scale $L_{\mathrm{ref}}=L_2\approx 1.0 \ \mathrm{fm}$. Other contributions arising from CLS ensembles are by far subleading at present. }
	\label{fig:mb_comparison}
\end{figure}

We also quote the b-quark mass in the $\overline{\mathrm{MS}}$ scheme in the $N_f=4$ theory, 
\begin{equation}
	\overline{m}_b^{(4)}(\overline{m}_b)=4.174(30)(12)_\Lambda \ \mathrm{GeV},
\end{equation}
where the second error  arises from $\Lambda^{(3)}_{\overline{\mathrm{MS}}}=341(12) \ \mathrm{MeV}$ \cite{PhysRevLett.119.102001}. We refer to Fig.~\ref{fig:mb_comparison} for a comparison of our result with other studies and for a summary of the error budget of the computation.

\section{$B^{(\star)}$ leptonic decays}
\begin{figure}[!h] 
	\centering
	\includegraphics[scale=0.46]{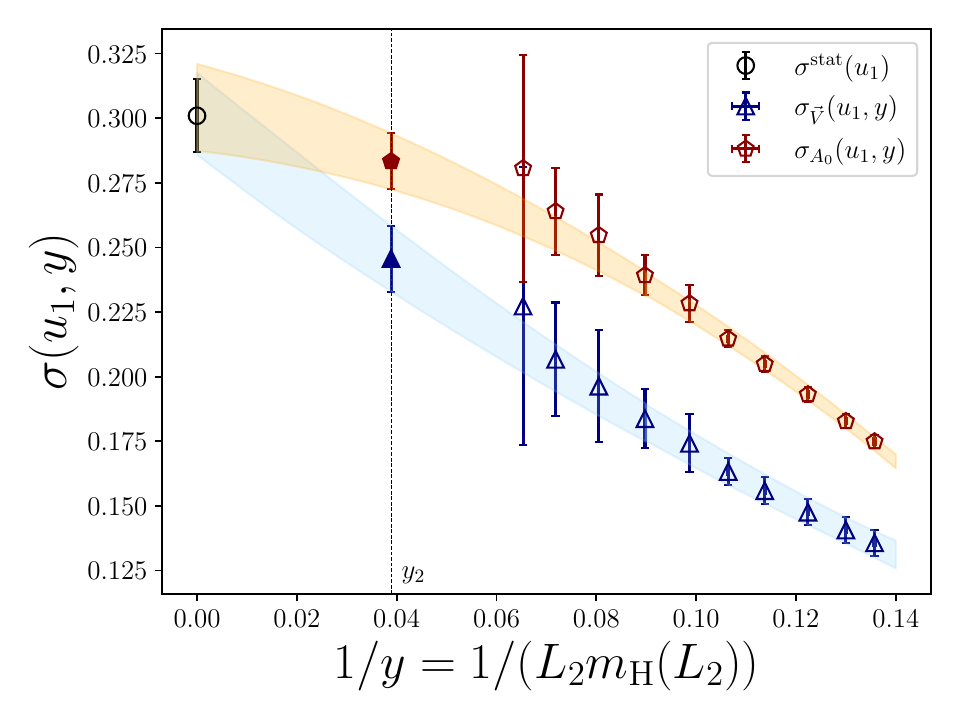}
	\includegraphics[scale=0.46]{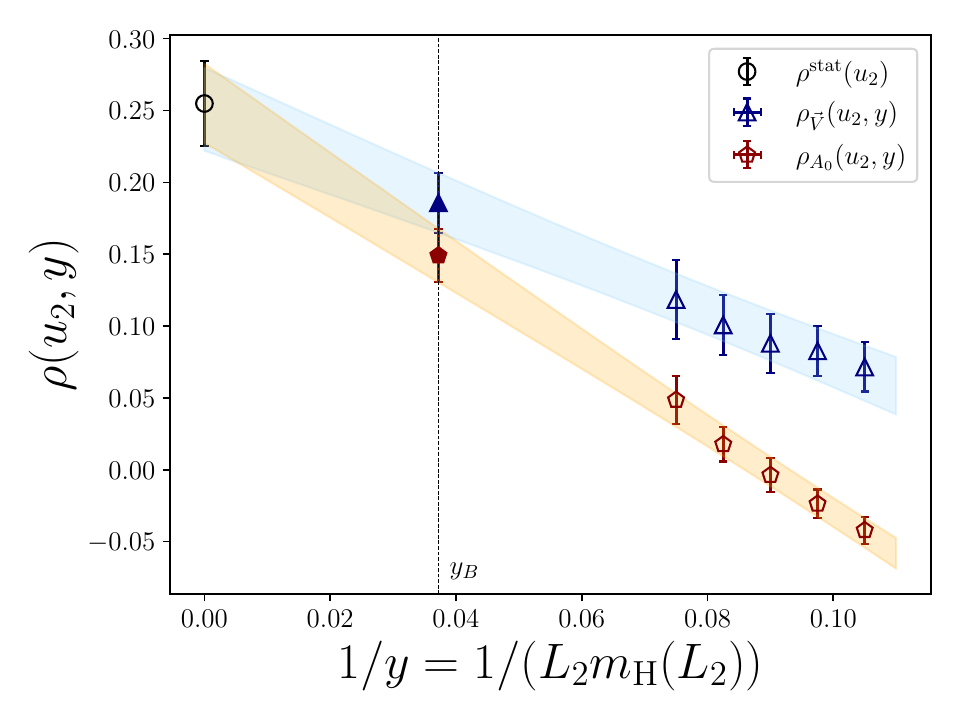}	
	\caption{Interpolation to the target b-quark scale for the step $L_1\to L_2$ (\textit{left}) and $L_2\to \mathrm{CLS}$ (\textit{right}). Blue triangles and red pentagons correspond to the vector and pseudo-scalar  meson decays  relativistic SSF, respectively, while the black circles are the shared static $\sigma^\mathrm{stat}$ and $\rho^{\mathrm{stat}}$.}
	\label{fig:ssf_decays}
\end{figure}

As discussed  in \cite{RainerPos}, our strategy is also suited for the extraction of  decay constants,  using the already determined quark mass proxies $y_i$. Here we present results for the vector meson decay constant $f_{B^\star}$, which in our strategy plays a crucial role in a precise determination of $b\to u$ semi-leptonic decays, and the pseudo-scalar $f_B$, used as a relevant crosscheck of our strategy. Given the axial-vector and vector heavy-light currents $A_\mu, V_\mu$, we define the  dimensionless observable used to build the SSF as
\begin{equation}
	\Phi_{A_0(\vec{V})}(u, y) = \ln\bigg(
	\frac{L_{\mathrm{ref}}^{3/2}}{\sqrt{2}} \hat{f}_{B^{(\star)}}	\bigg)
	, \qquad \hat{f}_{B^{(\star)}} = \sqrt{m_{B^{(\star)}}}f_{B^{(\star)}},
\end{equation}
where in the finite volume the matrix elements are constructed from correlation functions with Schr\"odinger functional boundary conditions \cite{Luscher:1992an, Sint:1993un,RainerPos}.  The step scaling chain  takes the form
\begin{eqnarray}
\ln\bigg(
\frac{L_{\mathrm{ref}}^{3/2}}{\sqrt{2}} \hat{f}_{B^{(\star)}}	\bigg) 
= 
\Phi_{A_0(\vec{V})}(u_1, y_1)
+
\sigma_{A_0(\vec{V})}(u_1, y_2)
+
\rho_{A_0(\vec{V})}(u_2, y_B),
\label{eq:decay_master_eq}
\end{eqnarray}
in terms of the SSFs
\begin{eqnarray}
	\sigma_{A_0(\vec{V})}(u_1, y) &=& \Phi_{A_0(\vec{V})}(u_2, y) -  \Phi_{A_0(\vec{V})}(u_1, y),
	\\
	 \rho_{A_0(\vec{V})}(u_2, y) &=& \Phi_{A_0(\vec{V})}( y) -  \Phi_{A_0(\vec{V})}(u_2, y).
\end{eqnarray}
Their lattice approximants $\Sigma,R$ use the same LCPs as before and we take their continuum limits in the exact same way. 
In Fig.~\ref{fig:ssf_decays} the interpolations for $\sigma_{A_0(\vec{V})}$ and $\rho_{A_0(\vec{V})}$ to the known $y_2$ and $y_B$ are shown, again with a second order polynomial in $1/y$. It is worth noting that in the $m_h\to \infty$ limit spin symmetry means that vector and pseudo-scalar share the same step scaling functions: $\sigma^{\mathrm{stat}}_{A_0}=\sigma^{\mathrm{stat}}_{\vec{V}}$ and $\rho^{\mathrm{stat}}_{A_0}=\rho^{\mathrm{stat}}_{\vec{V}}$.
For the flavour averaged combinations we arrive at
preliminary results
\begin{equation}
	f_{\overline{B}} = 205.4(4.7)\ \mathrm{MeV} \ [2.2\%], \qquad f_{\overline{B}^\star} = 207.0(5.4) \ \mathrm{MeV}\ [2.6\%].
\end{equation}
We observe that all the steps involved in the computation contribute significantly to the $f_{\overline{B}^{(\star)}}$ error budget, and increased precision can be reached with additional statistics. In contrast to the quark mass, now also our 200\% error for $c_A^\mathrm{stat}$ plays a significant role.
 
\begin{figure}[!h] 
	\centering
	\includegraphics[scale=0.46]{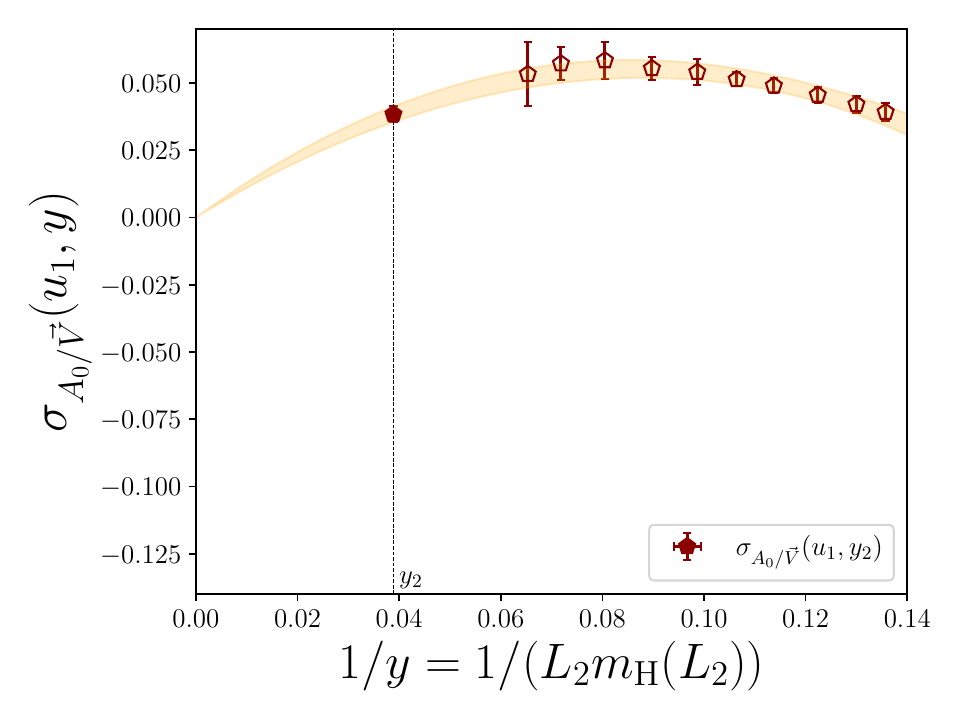}
	\includegraphics[scale=0.46]{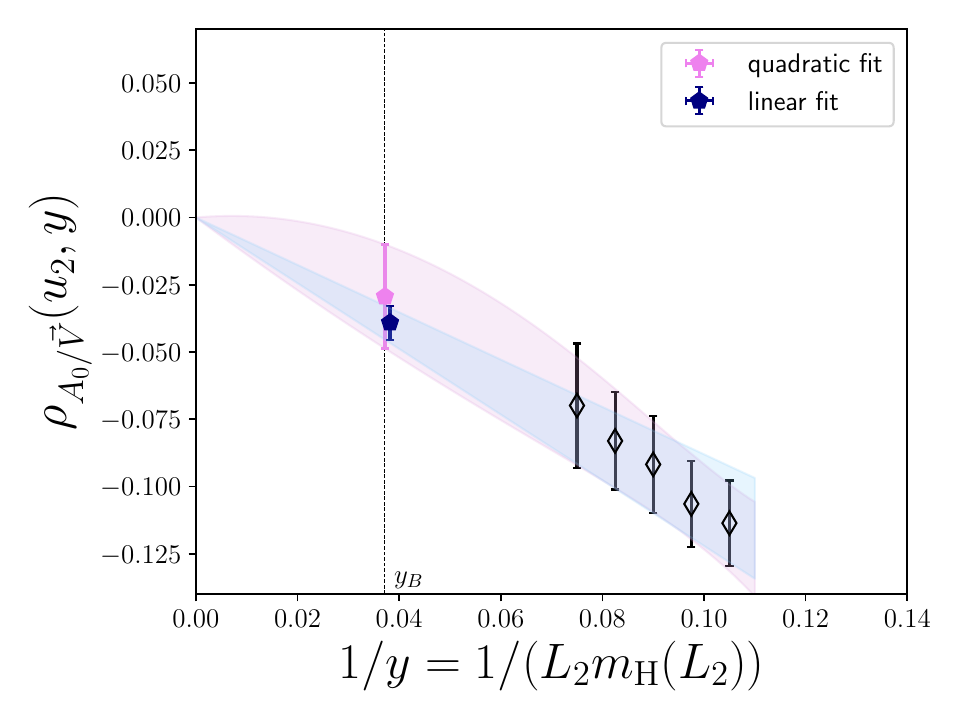}	
	\caption{Interpolation of the ratio $f_{\overline{B}}/f_{{\overline{B}}^\star}$ to the target b-quark scale for the steps $L_1\to L_2$ (\textit{left}) and $L_2\to \mathrm{CLS}$ (\textit{right}). On the right plot we show  results from both a linear and quadratic fit. We use the latter in our determination as conservative choice.}
	\label{fig:ssf_ratios}
\end{figure}

\begin{figure}[!h] 
	\centering
	\includegraphics[scale=0.6]{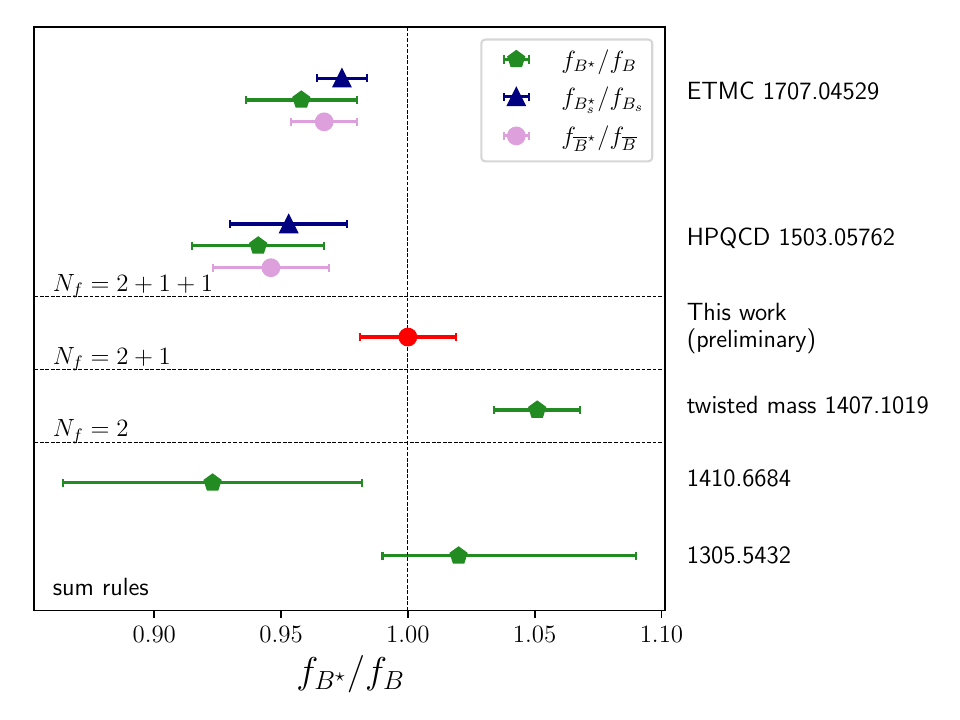}
		\caption{Results comparison for the ratio between vector and pseudo-scalar meson decays. Blue triangles correspond to $f_{B_s^{\star}}/f_{B_s}$, green pentagons to $f_{B^{\star}}/f_{B}$, while pink circles show the flavour averaged combination of the ratio $f_{{\overline{B}}^\star}/f_{\overline{B}}$. Starting from the bottom, results are taken from: \cite{Gelhausen:2013wia, Lucha:2014nba, Becirevic:2014kaa, Colquhoun:2015oha, Lubicz:2017asp}.   }
	\label{fig:ratios_comparison}
\end{figure}

An additional interesting quantity that we can extract with our strategy is the ratio $f_{{\overline{B}}^\star}/f_{\overline{B}}$. The latter is obtained from  the step scaling chain 
\begin{equation}
	\ln\bigg(
	\frac{\hat{f}_{\overline{B}}}{\hat{f}_{{\overline{B}}^\star}}	\bigg) 
	= 
	\Phi_{A_0/\vec{V}}(u_1, y_1)
	+
	\sigma_{A_0/\vec{V}}(u_1, y_2)
	+
	\rho_{A_0/\vec{V}}(u_2, y_B),
	\label{eq:ratio_master_eq}
\end{equation}
where we use the ratio between pseudo-scalar and vector matrix elements and $\sigma_{A_0/\vec{V}}=\sigma_{A_0}-\sigma_{\vec{V}}$, etc. We report  the interpolation results to the target $y_i$ in Fig.~\ref{fig:ssf_ratios}. Here we have  $\rho_{A_0/\vec{V}}^{\mathrm{stat}}=\sigma_{A_0/\vec{V}}^{\mathrm{stat}}=0$,  imposing a strong constraint on the interpolation formula.
Our preliminary result, 
\begin{equation}
	f_{\overline{B}^\star} / f_{\overline{B}} = 0.999(19)\ [1.9\%]\,,
\end{equation}
happens to be in agreement with the static value of one, but
it is derived from individual pieces with few percent $1/m$ corrections which cancel only in the final quantity.
A comparison with other determinations is shown in Fig.~\ref{fig:ratios_comparison}.

\section{Conclusion and Outlook}
We have demonstrated the applicability of the strategy presented in \cite{RainerPos} by combining static and relativistic computations in the continuum to reach the physical b-quark mass by interpolation. We  presented preliminary results for the RGI b-quark mass, pseudo-scalar and vector meson decay constants, demonstrating the potential of this approach to achieve high-precision results with minimal assumptions.

Looking ahead, the future trajectory of this research involves extending computations to lighter pion mass ensembles, to reach the physical point also in this variable. Furthermore, ongoing efforts are directed towards determining the $c_A^{\mathrm{stat}}$ improvement  coefficient for the L\"uscher-Weisz gauge action. Studies of systematic effects related to the variations of the functional forms used for the continuum-limit extrapolation \cite{Husung:2019ytz,Husung:2021mfl,Husung:2022kvi} and for the interpolation to the target b-quark scale, as well as different definitions of the finite volume observables, will be considered. 

We note that the result for the vector meson decay constant can be employed for a precise determination  of the $b\to u$ semi-leptonic form factor $f_\perp$ as pointed out in \cite{RainerPos}. All finite volume step scaling quantities are extrapolated to the continuum limit and do not need to be repeated with discretisations used for the form factor computations. An accurate knowledge of these quantities is required to deepen our knowledge of the SM  and to test for  new physics phenomena.

\begin{small}
	\section*{Acknowledgments}
	\noindent
	We thank Oliver B\"ar, Alexander Broll and Andreas J\"uttner for discussions.
	We acknowledge support from the EU projects EuroPLEx
	H2020-MSCAITN-2018-813942 (under grant agreement No.~813942),
	STRONG-2020 (No.~824093) and HiCoLat (No.~101106243),
	as well as from grants,
	PGC2018-094857-B-I00, PID2021-127526NB-I00, SEV-2016-0597,
	CEX2020-001007-S funded by MCIN/AEI,
	the Excellence Initiative of Aix-Marseille University - A*Midex
	(AMX-18-ACE-005),
	and by DFG, through the
	Research Training Group \textit{``GRK 2149: Strong and Weak
		Interactions -- from Hadrons to Dark Matter''} and the project 
	``Rethinking Quantum Field Theory'' (No. 417533893/GRK2575).
	J.H. wishes to thank the Yukawa Institute for
	Theoretical Physics, Kyoto University, for its hospitality.
	The authors gratefully acknowledge the Gauss Centre for Supercomputing e.V. (\url{www.gauss-centre.eu}) for funding this project by providing computing time on the GCS Supercomputer SuperMUC-NG at Leibniz Supercomputing Centre (\url{www.lrz.de}). We furthermore acknowledge the computer resources provided by the CIT of the University of Münster (PALMA-II HPC cluster) and by DESY Zeuthen (PAX cluster) and thank the staff of the computing centers for their support. We are grateful to our colleagues in the CLS initiative for producing the large-volume gauge field configuration ensembles used in this study.

\end{small}

\bibliographystyle{JHEP}
\bibliography{proceedings.bib}

\end{document}